\documentclass[aps,prl,twocolumn,titlepage,nofootinbib,longbibliography]{revtex4}
\usepackage{graphicx,physics}
\usepackage{color}
\usepackage{dcolumn}
\usepackage{latexsym}

\usepackage{hyperref,amssymb}
\usepackage{url}
\usepackage{color,soul}
\usepackage{graphicx}
\usepackage{multirow}
\usepackage{amsmath}
\newcommand{\beq}{\begin{eqnarray}}
\newcommand{\eeq}{\end{eqnarray}}
\usepackage{mathrsfs}
\usepackage{float,soul}
\usepackage[dvipsnames]{xcolor}
\usepackage{mathtools}
\usepackage{slashed}
\usepackage{physics}	
\usepackage{graphicx}   
\usepackage{epstopdf}
\usepackage{subfigure}  
\usepackage{hyperref}   
\usepackage{bbold}
\usepackage{wasysym}
\usepackage{feynmp}
\usepackage{hyperref}
\hypersetup{colorlinks,
}

\usepackage{tikz}
\newcommand{\redc}[2][red,fill=red]{\tikz[baseline=-0.5ex]\draw[#1,radius=#2] (0,0) circle ;}%
\newcommand{\bluec}[2][blue,fill=blue]{\tikz[baseline=-0.5ex]\draw[#1,radius=#2] (0,0) circle ;}

\usepackage[latin1]{inputenc}

\begin{document}

\title{\Large A geometric approach to predicting plasticity in disordered solids}
\author{Long-Zhou Huang$^{1,2}$}
\author{Xu Yang$^{3,4,5}$}
\author{Min-Qiang Jiang$^{1,2}$}
\email{mqjiang@imech.ac.cn}
\author{Yun-Jiang Wang$^{1,2}$}
\email{yjwang@imech.ac.cn}
\author{Matteo Baggioli$^{3,4,5}$}
\email{b.matteo@sjtu.edu.cn}
\affiliation{$^1$State Key Laboratory of Nonlinear Mechanics, Institute of Mechanics, Chinese Academy of Sciences, Beijing 100190, China}
\affiliation{$^2$School of Engineering Science, University of Chinese Academy of Sciences, Beijing 100049, China}
\affiliation{$^3$School of Physics and Astronomy, Shanghai Jiao Tong University, Shanghai 200240, China}
\affiliation{$^4$Wilczek Quantum Center, Shanghai Jiao Tong University, Shanghai 200240, China}
\affiliation{$^5$Shanghai Research Center for Quantum Sciences, Shanghai 201315, China}

\begin{abstract}
It was recently shown that vortex-like topological defects with negative winding number in the vibrational modes of a two-dimensional glass under quasistatic shear correlate strongly with plastic events, offering a promising route to predict them. However, many of these vortices, a number that actually grows quadratically with mode frequency, are entirely unrelated to plasticity and arise simply from the underlying plane-wave structure of the modes. This raises doubts about the fundamental relevance of such defects to plastic rearrangements and limits their predictive power. Here, we introduce a geometrical filter based on the Nye dislocation density that, when applied to the vibrational modes, removes these spurious defects and reveals the true plastic precursors. Using simulations of a two-dimensional model glass, we show that this filtered approach consistently outperforms the conventional vortex-based method, particularly at small strains and when focusing on genuine plastic stress drops, offering a more robust tool to predicting plasticity in glasses from their undeformed initial state.
\end{abstract}
\maketitle
\color{blue}\textit{Introduction} 
\color{black} -- \textit{Plasticity} refers to the ability of a material to undergo irreversible mechanical deformation without breaking, and it can be readily experienced in everyday life by bending a metal spoon or crushing a soda can. In crystalline materials, such as common metals, the microscopic origin of plasticity is well established and attributed to the behavior of topological structural defects known as \textit{dislocations} \cite{schmid1968plasticity,sutton2020physics}.

In contrast, in amorphous solids, which lack translational symmetry and a periodic lattice structure, yet still exhibit plasticity across a wide range of time and length scales \cite{Berthier2025}, the physical origin of plastic deformation remains one of the most fundamental open problems. The absence of long-range order and the apparently featureless structure prevents the identification of plastic defects directly from structural information, while also making formal approaches based on topology or geometry inapplicable without significant generalization.

Building on the pioneering work of Argon \cite{ARGON197947}, and later Falk and Langer \cite{PhysRevE.57.7192,LANGER2006375}, it is now established that plasticity in amorphous solids originates from localized regions involving roughly $10$-$100$ particles, which undergo highly non-affine rearrangements. These localized defective clusters, commonly referred to as \textit{shear transformation zones} (STZs) or, more broadly, \textit{soft spots}, form the conceptual foundation of the most widely used elastoplastic mesoscopic models of mechanical response in glasses \cite{annurev:/content/journals/10.1146/annurev-conmatphys-062910-140452}.

Following the original proposal in \cite{PhysRevE.57.7192}, the definition and identification of plastic spots in glasses typically rely on the kinematic measure $D^2_{\text{min}}$, which quantifies the deviation from local elastic and affine deformation. Although widely used and highly effective, this method depends on the particle displacement field obtained \emph{after} deformation and therefore does not enable an \emph{a priori} prediction of plastic activity from the initial, undeformed configuration. In this sense, it serves more as a dynamical diagnostic than a predictive tool. Although several structural indicators have been proposed to identify plastic-prone regions in glasses \cite{PhysRevMaterials.4.113609}, a robust and fundamental understanding remains elusive.

In 2023, Wu et al.\ \cite{wu2023topology,Baggioli2023} proposed that topology, and in particular vortex-like topological defects, could still serve as a fundamental building block and predictive tool for plasticity in glasses, provided topological analysis is applied not to the atomic structure itself but to the vibrational modes of the \emph{initial} configuration. Using a two-dimensional (2D) simulated glass, they showed that the spatial organization of antivortex defects in the low-frequency vibrational modes exhibits a strong correlation with the plastic spots emerging under shear, thereby offering a  method for predicting their locations directly from the undeformed structure.

These findings were quickly confirmed experimentally in a 2D colloidal glass \cite{Vaibhav2025} and subsequently generalized to three-dimensional systems \cite{10.1093/pnasnexus/pgae315,bera2025hedgehogtopologicaldefects3d,wu2024geometrytopologicaldefectsglasses}. Furthermore, it was demonstrated that the same methodology, when applied to crystalline materials, correctly identifies their intrinsic topological structural defects, including dislocations \cite{HUANG2025106274}.

Despite the unquestionable success of this method, it was soon recognized, in fact by Wu et al.\ themselves \cite{wu2023topology} (see Supplementary Fig.~4 therein), that the eigenvector field of pure plane-wave solutions is riddled with vortex-like topological defects. These vortices are not associated with plasticity but arise generically from the nodal structure of plane waves, forming highly ordered patterns with a characteristic spacing set by the wavelength. This is also reflected in their number scaling with frequency in the low-frequency regime obtained in simulations \cite{wu2023topology}.

\begin{figure}[ht]
    \centering
    \includegraphics[width=\linewidth]{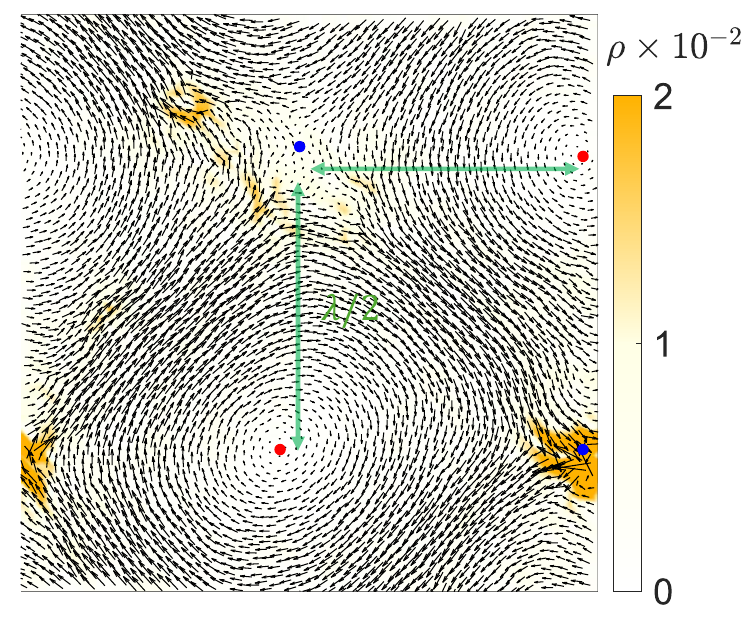}
\caption{\textbf{Filtering out spurious defects in the eigenvector field of a simulated glass.} Example of an eigenvector field in which \bluec{3pt} and \redc{3pt} denote, respectively, the $-1$ and $+1$ topological vortex defects defined in \cite{wu2023topology}. 
The background colormap displays the norm $\rho$ of the continuous Nye vector, Eq.~\eqref{nye}. 
The green vectors indicate half of the phonon wavelength associated with this eigenvector.
}
    \label{fig:1}
\end{figure}

More generally, this observation casts doubts on whether these vortex defects, despite their apparent correlation with plasticity, are truly the fundamental carriers of plastic deformation. Taken individually, and different from dislocations in crystals, they clearly cannot be. 

The underlying issue lies in the mathematical definition of the vortex defects themselves. An eigenvector field is intrinsically a vector field, and its singularities cannot be characterized solely through the directional angle $\theta$. Specifically, for a 2D vector $\vec{v} = (v_x, v_y)$, the phase defined through $\theta \equiv \arctan(v_y / v_x)$ is, by construction, ill-defined at any point where $|\vec{v}| = 0$. These apparent singularities are spurious artifacts of the polar representation, not genuine defects of the vector field. 

Filtering out these spurious vortex-like defects and identifying the genuine defects in the eigenvector fields that are responsible for plasticity in glasses are the central questions addressed in this work.

\color{blue}{\it Filtering method} \color{black} -- Consider the two-dimensional eigenvector field $\vec{e}^{\,\omega}(x,y)$ with frequency $\omega$. Following the original proposal of Ref. \cite{wu2023topology}, vortex defects can be identified through the winding number $q$,
\begin{equation}
    q = \frac{1}{2\pi} \oint d\theta, 
    \qquad 
    \theta \equiv \arctan\!\left(\frac{e^{\,\omega}_y}{e^{\,\omega}_x}\right).\label{eq1}
\end{equation}
Vortices and antivortices correspond to singularities of the phase field with charges $q=\pm 1$.

In contrast, we apply the Nye operator to the eigenvector field and define the corresponding continuous Nye density as
\begin{equation}
    \alpha_i(x,y) = \epsilon_{3lk}\,\partial_l \partial_k\, e_i(x,y), 
    \label{nye}
\end{equation}
where $\epsilon$ is Levi-Civita symbol and the index $3$ denotes the out-of-plane direction perpendicular to the 2D system. We indicate its norm as $\rho \equiv |\vec{\alpha}|$. When the eigenvector field is replaced by the particle displacement field, $\alpha_i$ reduces to the continuous Nye dislocation density or Burgers-vector density familiar from crystalline elasticity \cite{NYE1953153,kleinert1989gauge}. Physically, it measures the density of translational singularities in the corresponding vector field, analogous to dislocations in crystals. Geometrically, it is a measure of torsion \cite{kleinert1989gauge,Kupferman2015}.

When applied to discrete numerical data, however, the Nye tensor cannot distinguish genuine topological singularities from \textit{continuous defects} \cite{RevModPhys.80.61}, which act as sources of geometric incompatibility. For this reason, in the present context $\alpha_i$ does not measure topological singularities but instead quantifies geometric incompatibility, corresponding to violations of the integrability condition for the underlying vector field. Bona-fide topological defects would correspond to quantized delta-function singularities in $\vec{\alpha}$, while continuous and non-quantized values correspond to geometric singularities.

Importantly, as demonstrated in Fig.~\ref{fig:5} in the \textit{End Matter}, either a crystalline dislocation or Eshelby inclusion produces a clear and localized signal in $\rho$. In contrast, ideal vortices and antivortices yield a vanishing value of $\rho$. This key property makes $\rho$ an effective tool for filtering out spurious defects that do not correspond to genuine geometric singularities of the vector field but instead arise solely from singularities of its phase.

\begin{figure}[ht]
    \centering
    \includegraphics[width=\linewidth]{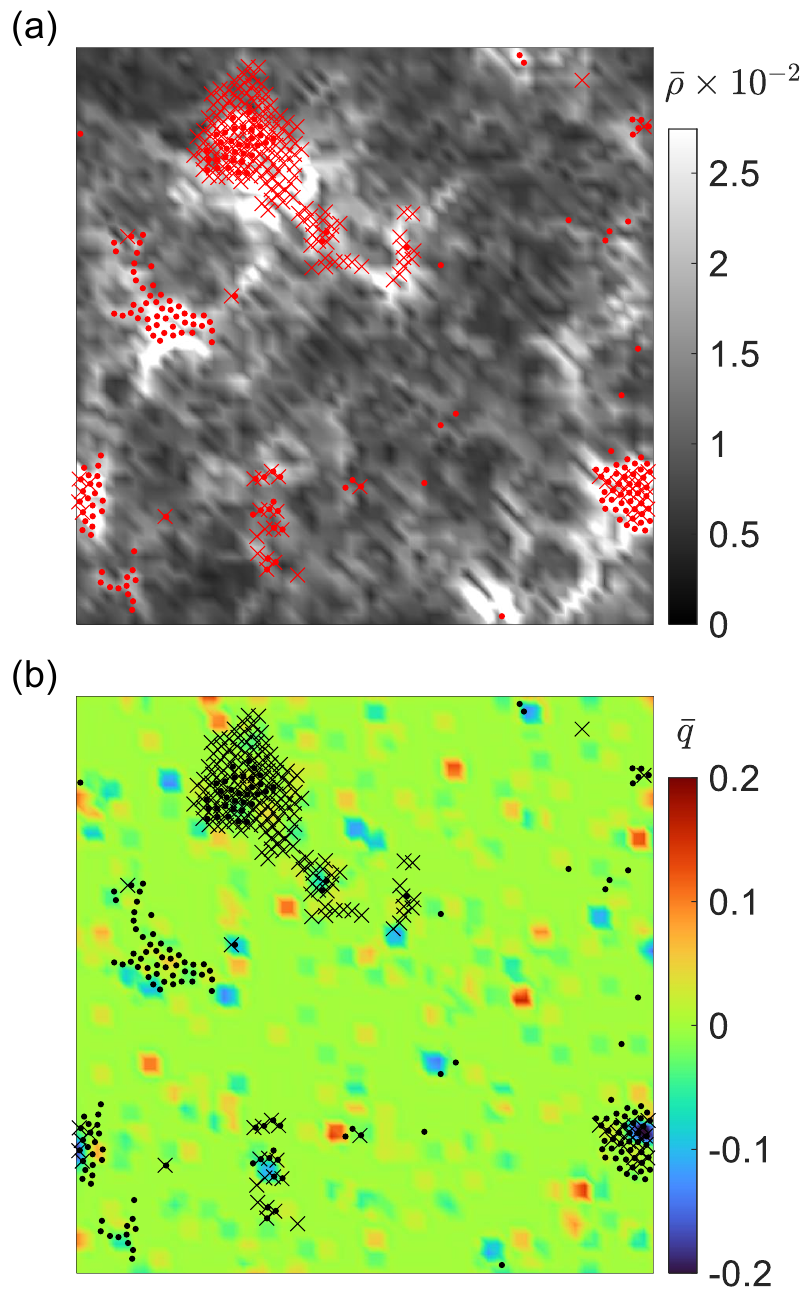}
    \caption{\textbf{Geometric Nye density versus average winding number.} \textbf{(a)} Average local Nye density $\bar{\rho}(x,y)$, Eq. ~\eqref{localnye}, corresponding to the first $n=20$ vibrational modes. Red crosses and disks correspond to the plastic spots, identified using $D^2_{\text{min}}$, for a small strain $\gamma=0.005$ for direct and inverse shear. \textbf{(b)} Same representation where the color map indicates the average value of the topological winding number $\bar{q}(x,y)$.}
    \label{fig:2}
\end{figure}

\begin{figure*}[ht]
    \centering
    \includegraphics[width=\linewidth]{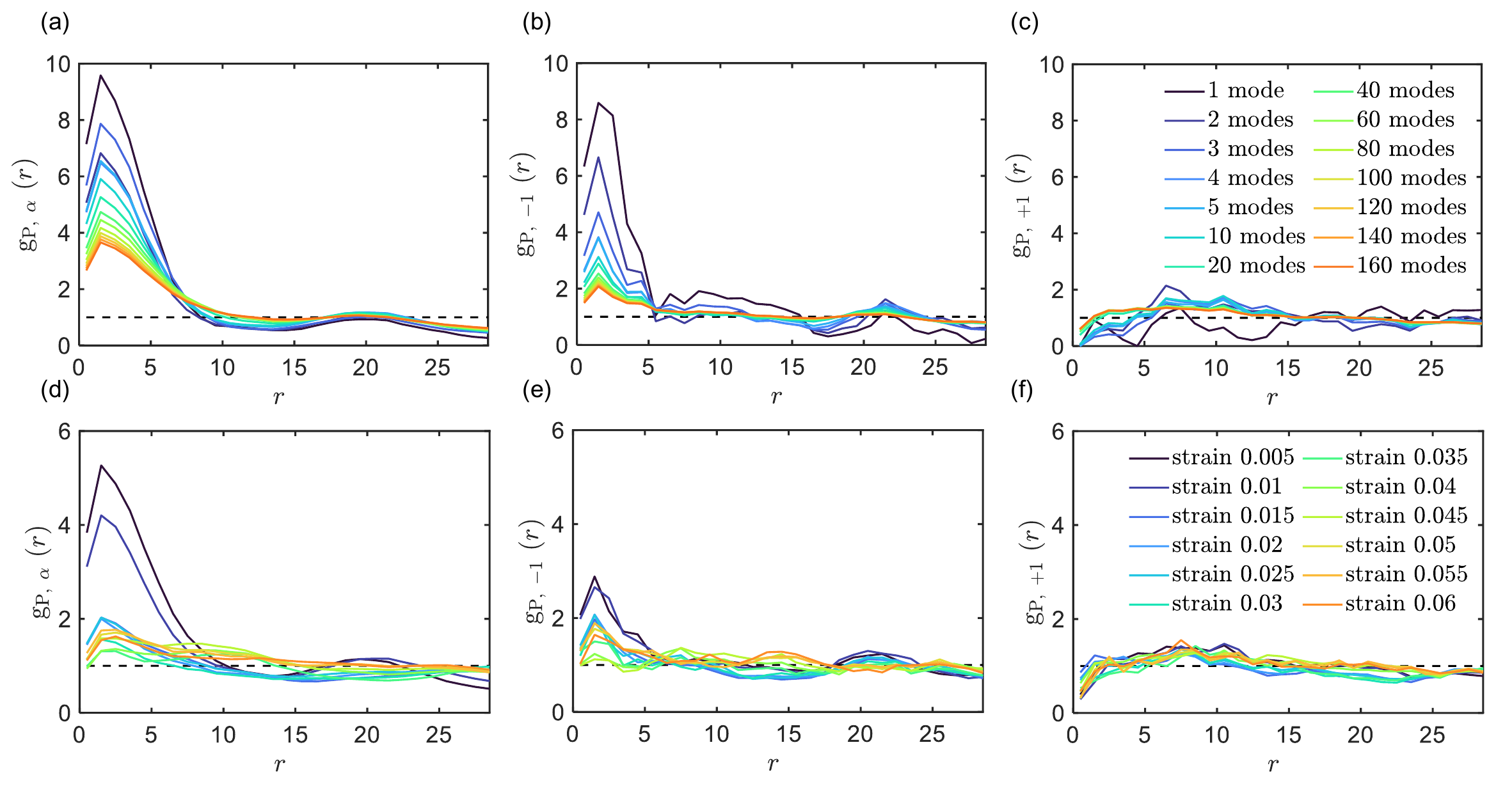}
    \caption{\textbf{Correlation with plastic soft spots.} \textbf{(a)-(c)} Radial pair correlation functions between plastic spots (P), Nye density $\alpha$, and $\pm 1$ topological defects. Different colors correspond to different modes involved in the average. The global strain value is fixed to $\gamma=0.005$. \textbf{(d)-(f)} Same analysis for fixed value of $n=20$ and different values of the strain used to compute the non-affine $D^2_{\text{min}}$ measure.}
    \label{fig:3}
\end{figure*}

To illustrate this filtering capability, Fig.~\ref{fig:1} shows the spatial structure of an eigenvector field obtained from simulations of a simulated glass, corresponding to the lowest frequency $\omega = 0.38$ (more examples are provided in the Supplementary Information, SI). Details of the simulated model and numerical methods are provided in the \textit{End Matter} and SI. The eigenvector field in Fig.~\ref{fig:1} contains two vortices (\redc{3pt}) and two antivortices (\bluec{3pt}), identified using the local winding number defined in Eq.~\eqref{eq1}. These four phase defects do not correspond to singularities of the vector field itself; they are simply features inherited from the underlying plane-wave structure. Indeed, as emphasized by the green arrows, their ordered arrangement at a separation of $\lambda/2=27.4$ (with $\lambda=2\pi c_{\mathrm{T}}/\omega$ the phonon wavelength and $c_{\mathrm{T}}=3.34$ the transverse wave velocity) is precisely what one expects for a plane wave of the form $\vec{e} = (\sin(2\pi y/\lambda),\, \sin(2\pi x/\lambda))$.

By contrast, when we apply the Nye density to the same eigenvector field, new features emerge that cannot be inferred from the vortex-like defects. As expected (see zoomed view presented in the \textit{End Matter}), the Nye density $\rho$ highlights localized regions in which the eigenvector field exhibits a disordered and singular behavior consistent with translational incompatibility.

\color{blue}{\it Demonstration of its performance} \color{black} -- We consider a 2D Kob-Andersen model glass under athermal quasi-static simple shear (AQS) along the $xy$ or $-xy$ plane \cite{PhysRevE.74.016118} (see \textit{End Matter} and SI for details). We first consider a small shear strain $\gamma=0.005$ and identify plastic spots using particles among the top $5\%$ in the non-affine measure $D^2_{\text{min}}$ \cite{PhysRevE.57.7192}. These soft spots are depicted with crosses and disk symbols in panels (a)-(b) of Fig.~\ref{fig:2}, corresponding respectively to direct $xy$ and inverse ($-xy$) shear. The color map in panel (a) of Fig.~\ref{fig:2} represents the weighted average value of the Nye density,
\begin{equation}
    \bar{\rho}\equiv \frac{1}{n}\sum_{i=1}^n \frac{\rho_i(x,y)}{\omega_i^2},\label{localnye}
\end{equation}
computed over the first $n=20$ modes. Different choices of the weight function are explored in the SI and yield similar results.

\begin{figure*}
    \centering
    \includegraphics[width=\linewidth]{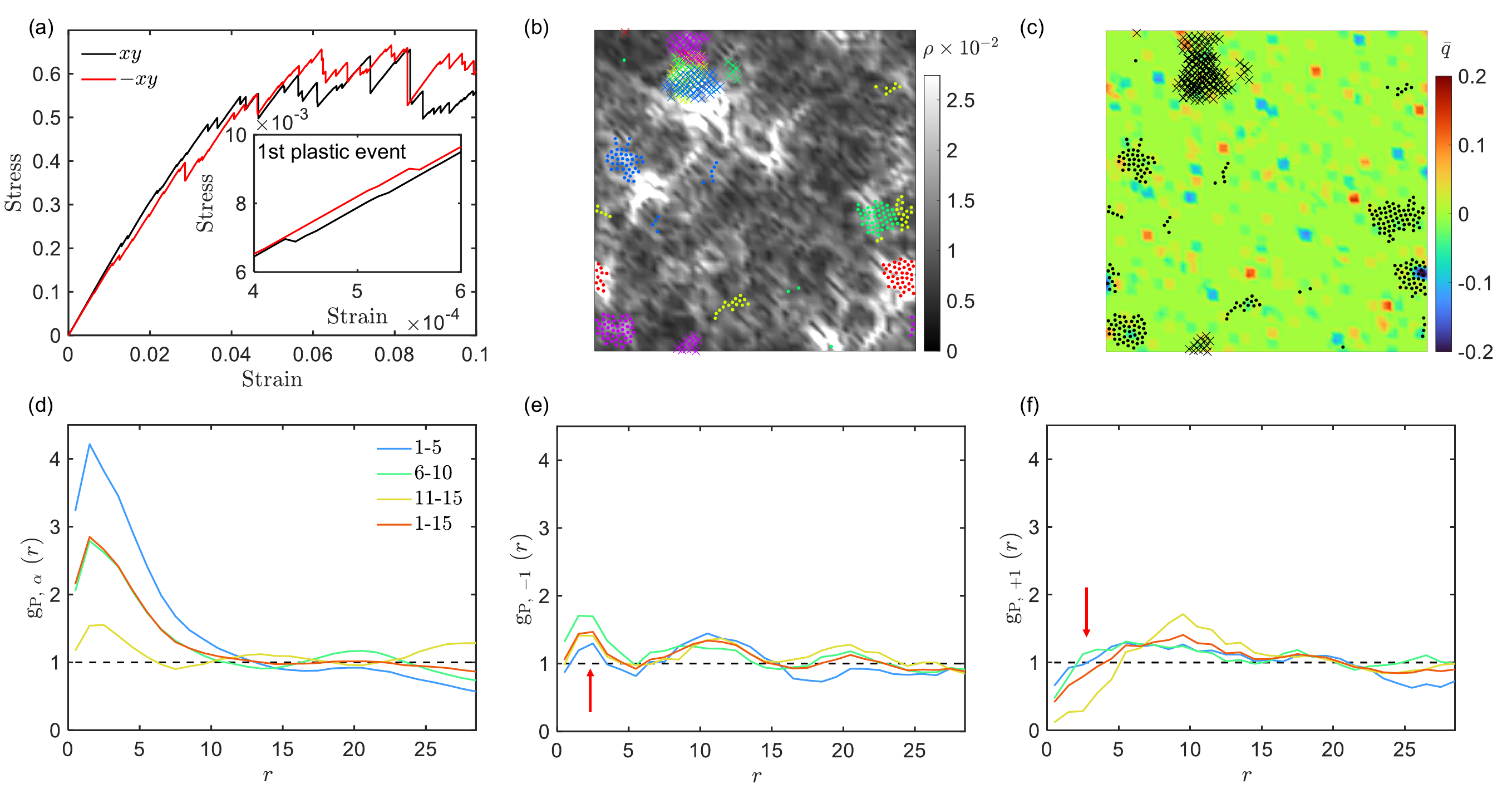}
    \caption{\textbf{Identifying the real plastic events.} \textbf{(a)} Stress-strain curve under shear and zoom on the first plastic event. Red and black curves correspond respectively to $xy$ and inverse ($-xy$) shear. \textbf{(b)} Colored clusters are the plastic spots corresponding to the first five plastic events (PEs) in the stess-strain curve. The background color is the averaged Nye density computed using the first $n=20$ vibrational modes. \textbf{(c)} In black, the same plastic soft spots identified with $D^2_{\text{min}}$. The background color is the average winding number from the first $n=20$ modes. \textbf{(d)-(f)} Radial pair correlation functions between the plastic spots (P), the average Nye density ($\alpha$) and the $\pm 1$ topological charge. Different colors correspond to different PE considered as indicated in the legend. Red arrows in panel (e) and (f) indicate respectively a weak correlation/anti-correlation consistent with the results of \cite{wu2023topology}.}
    \label{fig:4}
\end{figure*}

The spatial correlation between the areas with large $\bar{\rho}$ and the plastic spots is evident by the naked eye. On the other hand, in panel (b) of Fig.~\ref{fig:2} we follow the method proposed by Wu et al.\ \cite{wu2023topology} and display the local value of the average winding density,
\begin{equation}
    \bar{q}=\frac{1}{n}\sum_{i=1}^n\frac{q_i(x,y)}{\omega_i^2},\label{wind}
\end{equation}
computed on the same number of modes. A weak correlation between the areas with negative charge and the plastic spots is observed, consistent with the results in \cite{wu2023topology}.

To quantify more precisely the correlation between these quantities, we compute the pair correlation function $g(r)$ between plastic events and defects; further details are provided in the SI. Panels (a)--(c) of Fig.~\ref{fig:3} show the correlation function between the averaged Nye density and the $\pm 1$ vortex topological defects with the plastic spots under shear. We perform this analysis by varying the number of modes included in Eqs.~\eqref{localnye}--\eqref{wind}, from $n=1$ (only the lowest-frequency mode) up to $n=160$, which corresponds approximately to the boson-peak frequency and the end of the Debye regime (see SI).

First, we reproduce the findings of \cite{wu2023topology}: $-1$ topological defects display a pronounced short-range correlation with plastic spots, whereas $+1$ defects exhibit a weak anti-correlation. More generally, we find that, regardless of the number of modes included in the averaging, the Nye measure consistently outperforms the topological winding-number measure, exhibiting a significantly stronger correlation with the plastic spots under shear deformation. This advantage becomes especially clear when many modes are included, a regime in which the correlation features associated with the topological charge become very weak or even disappear. We attribute this improvement to the ability of the Nye density to filter out spurious defects that bear no physical relation to plasticity, as illustrated in the simple example of Fig.~\ref{fig:1} (and demonstrated for more cases in the SI).

In panels (d)-(f) of Fig.~\ref{fig:3}, we further examine these correlations by varying the strain at which plastic spots are identified, effectively probing how far our plasticity predictors can reach. We find that, for small strains, the Nye-density method consistently outperforms the topological approach. However, at intermediate, though not very small, strain values, the two methods yield comparable (and weak) levels of correlation.

\color{blue}{\it Predicting the real plastic events} \color{black} -- Up to this point, we have followed the analysis of \cite{wu2023topology} by identifying local plastic spots at fixed values of the strain. In doing so, we have not distinguished between local rearrangements within the elastic regime and the \emph{global} plastic events that manifest as sudden stress drops in the stress--strain curve. We now shift our focus exclusively to these global plastic events, ignoring the piecewise elastic segments between them. For simplicity, and as illustrated in the stress--strain curve shown in panel (a) of Fig.~\ref{fig:4}, we concentrate on the first fifteen global plastic events within strain $\gamma \in [0,0.015]$, grouping them into three sets of five events in order to ensure sufficient statistical sampling. 

In panel (b) of Fig.~\ref{fig:4}, we show the locations of the soft spots associated with the first five plastic events (indicated by different colors), identified using the top $1.25\%$ of the $D^2_{\text{min}}$ measure. A clear correlation with the averaged Nye density computed from the first $n=20$ modes remains visible. Panel (c) displays the same soft spots superimposed on the averaged winding-number field. In this case, the correlation is visibly weaker.

Panels (d)--(f) of Fig.~\ref{fig:4} present the corresponding pair-correlation functions, following the same procedure as before. The four curves represent the correlations computed from the first five plastic events ($1$-$5$, light blue), the following five ($6$-$10$, light green), the next five ($11$-$15$, light yellow), and the full set of fifteen events ($1$-$15$, light orange). Notably, the correlation between the plastic spots and the $-1$ topological defects becomes very weak when only the first genuine plastic events are considered, and remains weak across all sets of events examined. This demonstrates that these defects are not robust predictors of plasticity.

In contrast, the clusters identified via the Nye geometric density continue to show a strong short-range correlation with the plastic soft spots, specially for the first low strain plastic events. This correlation weakens substantially only when the high-strain plastic events ($11$-$15$) are considered. However, even when all the first $15$ events are considered, the correlation of the Nye clusters is much stronger than the ones of the $-1$ vortices.

In the \textit{End Matter}, we have repeated this analysis for another independent sample. The results therein confirm this trend and demonstrate the robustness of our findings.


\color{blue}{\it Outlook} \color{black} -- In this work, we introduced the concept of continuous Nye density applied to the vibrational modes of amorphous solids. We demonstrated that this geometric measure effectively filters out vortex-like topological defects in the vibrational modes that are unrelated to plasticity and instead arise directly from the plane-wave character of the modes.

Applying this approach to a simulated glass under quasi-static shear, we showed that the continuous Nye density consistently outperforms the topology-based method of Wu et al.~\cite{wu2023topology} in predicting plastic soft spots. Furthermore, when focusing exclusively on the genuine plastic events identified as stress drops in the stress--strain curve, the method of Wu et al.\ loses predictive power and exhibits weak correlation with plasticity. In contrast, our geometric method remains effective at identifying plastic soft spots up to relatively large strains, including values approaching the yielding point.

Finally, we note that our geometric approach based on the Nye density extends naturally to three-dimensional systems. In this setting, it (i) dramatically reduces the computational cost associated with searching for microscopic topological defects, and (ii) circumvents the ambiguity of determining which type of defect, such as point defects \cite{bera2025hedgehogtopologicaldefects3d} or line defects \cite{wu2024geometrytopologicaldefectsglasses}, is most relevant for plasticity. Another interesting direction for future work is to combine our geometric approach with the vortex-based topological defects introduced in \cite{wu2023topology}, in order to identify which of those defects are genuinely relevant for plasticity.



\color{blue}{\it Acknowledgments} \color{black} -- We thank Zhenwei Wu, Peter Keim, Tim Petersen, Peter Harrowell, Walter Kob, Chengran Du, Amelia Liu, Jie Zhang, Alessio Zaccone, Ido Regev, Arabinda Bera and Wang Xin for useful discussions. We thank Zhenwei Wu for comments on a preliminary version of this manuscript.
YJW is supported by the National Key R\&D Program of China (No. 2025ZD0122000), the Strategic Priority Research Program of Chinese Academy of Sciences (No. XDB0620103 and No. XDB0510301), and the National Natural Science Foundation of China (Grant No. 12472112).
MQJ was supported by the National Outstanding Youth Science Fund Project (No. 12125206) and Major International Joint Research Projects (No. W2411003) of the National Natural Science Foundation of China.
MB acknowledges the support of the Shanghai Municipal Science and Technology Major Project (Grant No.2019SHZDZX01) and the support of the sponsorship from the Yangyang Development Fund.

\section{End Matter}\label{end}
\begin{figure}[h]
    \centering
    \includegraphics[width=\linewidth]{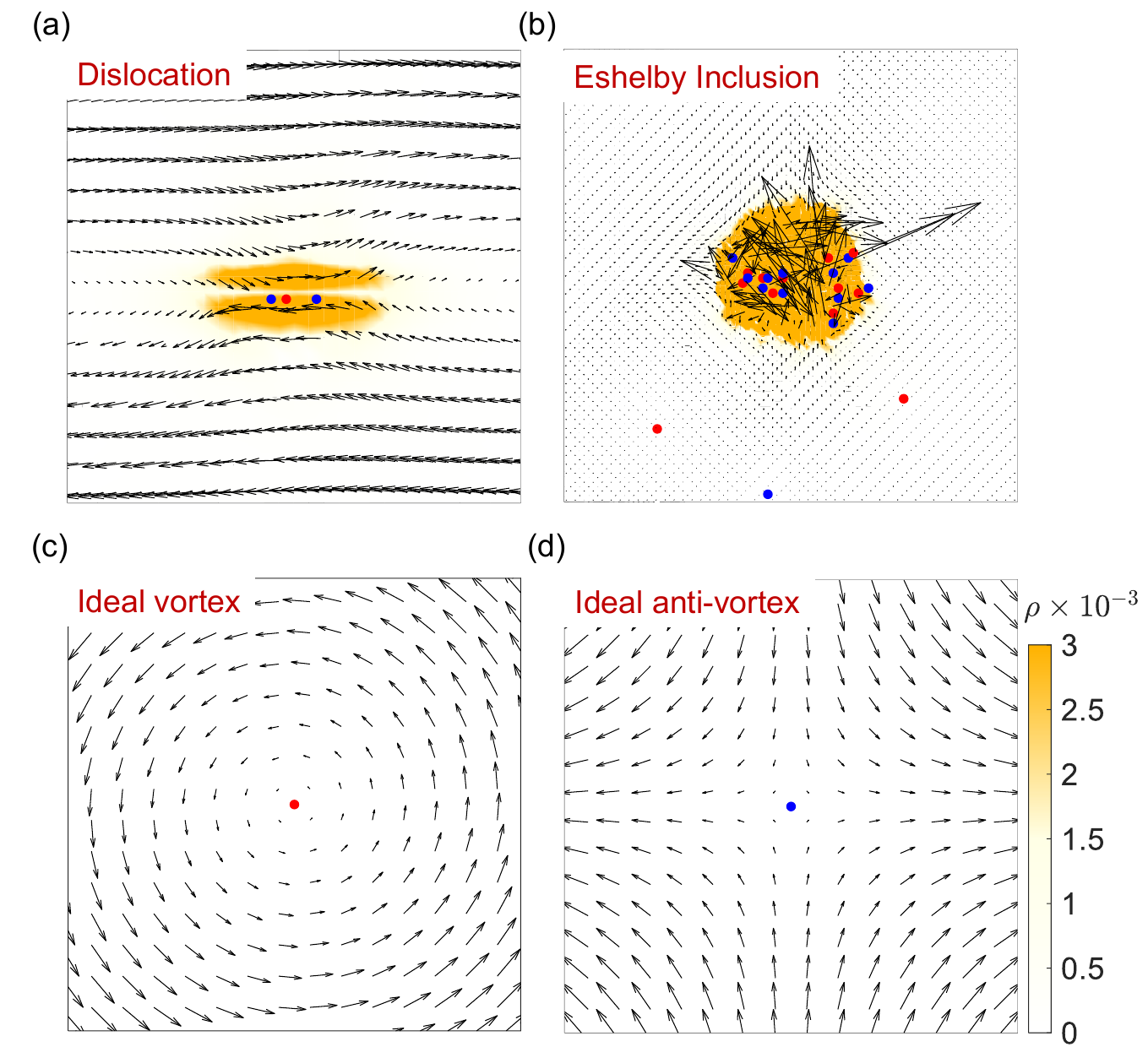}
    \caption{\textbf{Nye density for ideal cases.} \textbf{(a)} Lowest eigenvector field of an ideal dislocation \cite{HUANG2025106274}. \textbf{(b)} Lowest eigenvector field for an Eshelby inclusion \cite{HUANG2025106274}. \textbf{(c)} Eigenvector field for an ideal vortex. \textbf{(d)} Eigenvector field for an ideal anti-vortex. In all panels, the background color refer to the value of the local Nye density, while \bluec{3pt} and \redc{3pt} denote, respectively, the $-1$ and $+1$ topological vortex defects.}
    \label{fig:5}
\end{figure}
\subsection{Filtering out isolated vortices} 
Using the simulation methods described in Ref.~\cite{HUANG2025106274}, panels (a) and (b) of Fig.~\ref{fig:5} show that the Nye geometric density correctly identifies an edge dislocation in a BCC iron crystal and an Eshelby inclusion consisting of a soft BCC iron region embedded in a diamond lattice. In contrast, panels (c) and (d) apply the same geometric filter to the vector fields of an ideal vortex $\mathbf{U}(x,y)$ and an anti-vortex $\mathbf{V}(x,y)$, defined analytically as
\begin{equation}
    \mathbf{U}(x, y)=(-y, x), \qquad
    \mathbf{V}(x, y)=(x, -y).
\end{equation}
As shown, the Nye density vanishes identically in both cases.

\begin{figure}[ht]
    \centering
    \includegraphics[width=0.65\linewidth]{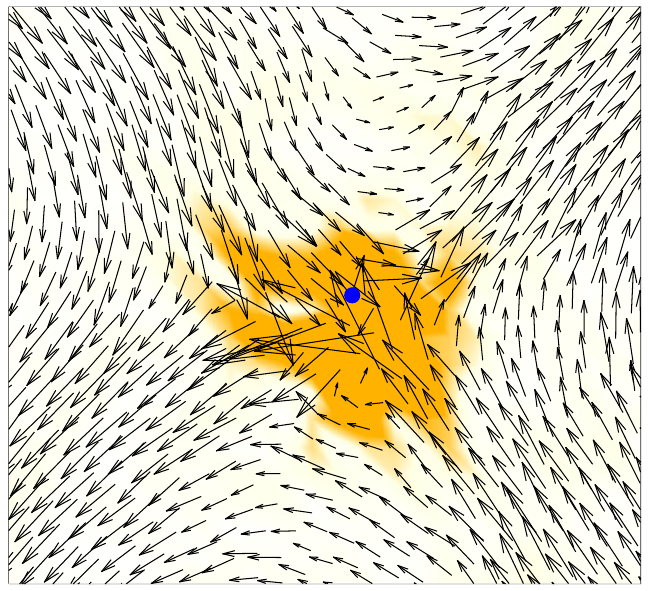}
    \caption{\textbf{Geometric incompatibility and Nye density.} A zoom of a highly incompatible region in the lowest eigenvector field presented in Fig.~\ref{fig:1} in the main text. The color map refers to the Nye density $\rho$. The blue symbol is an anti-vortex defect identified using the winding number.}
    \label{fig:6}
\end{figure}

\subsection{Regions with nonzero Nye density} 
To illustrate more clearly what the geometric Nye density captures, Fig.~\ref{fig:6} shows a zoomed view of an eigenvector field in a region where $\rho$ is nonzero. The presence of translational singularities in the vector field is evident.

\subsection{Validation of the results on a different independent sample}
\begin{figure}[ht]
    \centering
    \includegraphics[width=\linewidth]{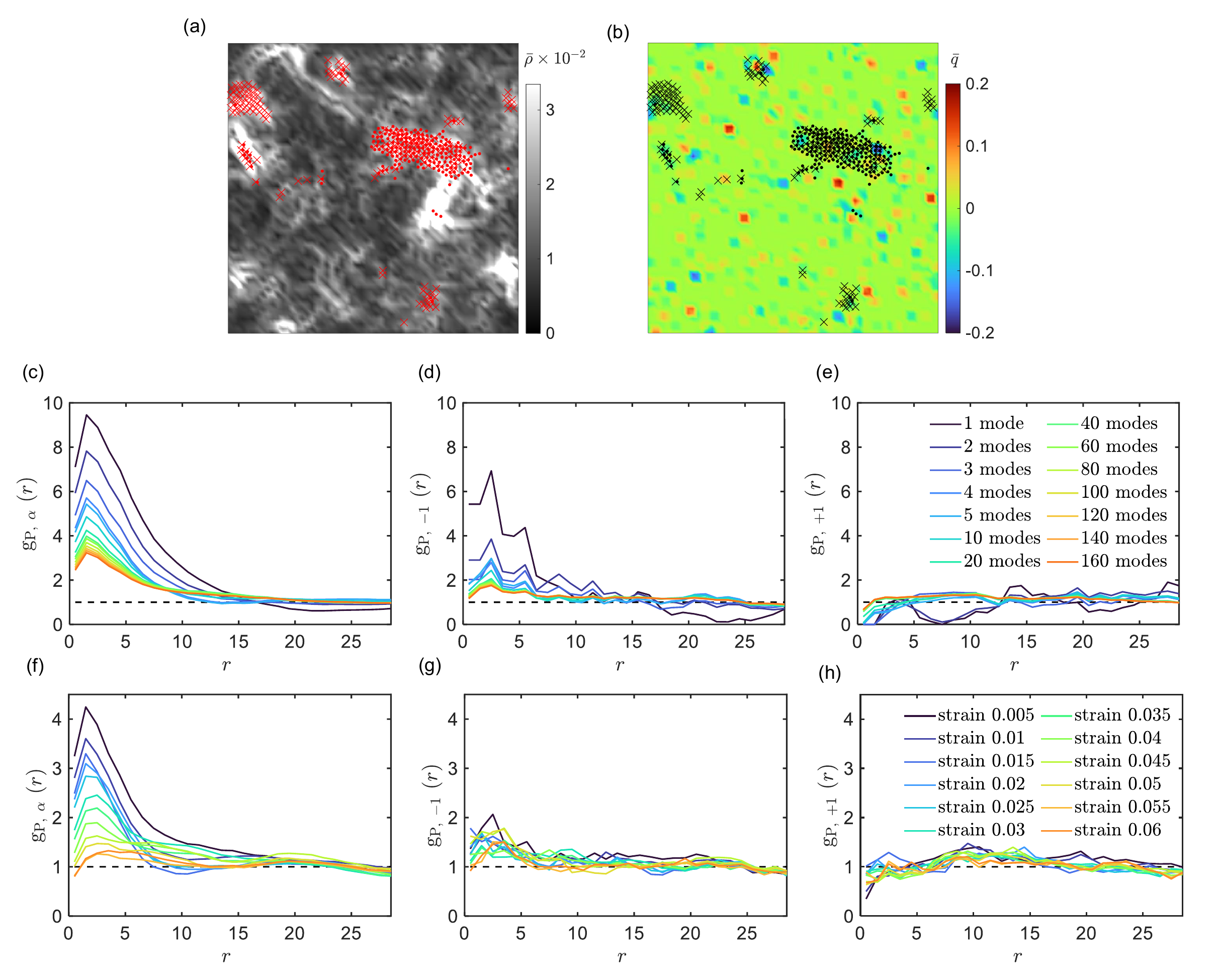}
    \caption{\textbf{Geometric approach versus topological densities.} Independent validation of the results presented in the main text in Figs. \ref{fig:2} and \ref{fig:3} using a different and independent sample. The superiority of the geometric approach with respect to the topological measure is evident.}
    \label{fig:s7}
\end{figure}

\begin{figure}[ht]
    \centering
    \includegraphics[width=\linewidth]{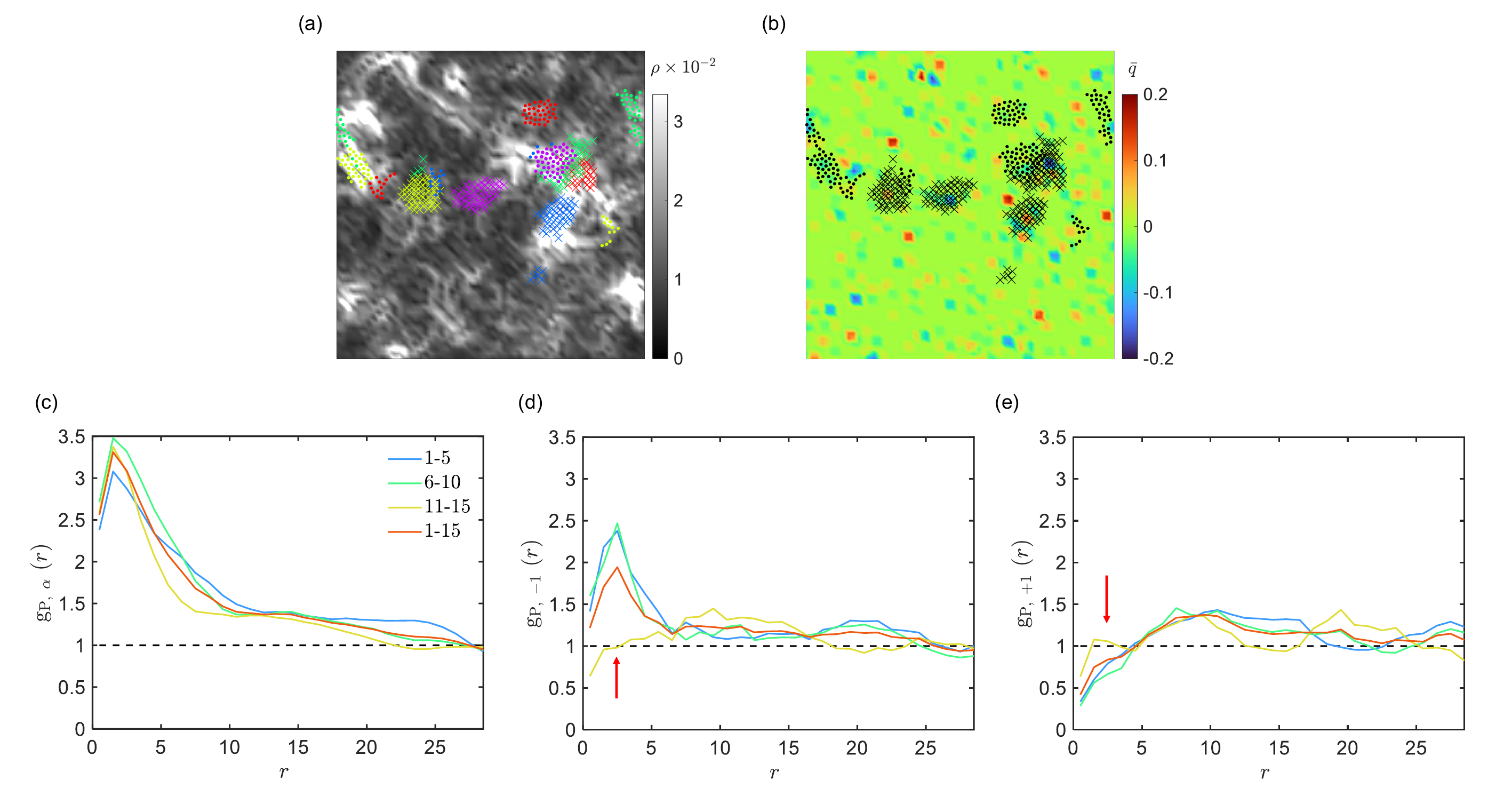}
    \caption{\textbf{Demonstration of the higher performance of the geometric approach.} Independent validation of the results presented in the main text in Fig. \ref{fig:4} using a different and independent sample. The superiority of the geometric approach with respect to the topological measure is evident.}
    \label{fig:s8}
\end{figure}

In order to confirm the universality of the results reported in the main text, we repeated the analysis using an independently prepared sample. All the main results are summarized in Figs. \ref{fig:s7} and \ref{fig:s8} and they confirm the outcomes reported in the main text.

\newpage
\onecolumngrid
\appendix 
\clearpage
\renewcommand\thefigure{S\arabic{figure}}    
\setcounter{figure}{0} 
\renewcommand{\theequation}{S\arabic{equation}}
\setcounter{equation}{0}
\renewcommand{\thesubsection}{SM\arabic{subsection}}
\section*{\Large Supplementary Material}
In this Supplementary Material, we provide further details about the numerical methods and the stability of our analysis. We also provide an additional analysis on an independent sample to demonstrate the universality of our findings.

\subsection{Simulation model}
The Kob--Andersen glass model was introduced by Kob and Andersen \cite{PhysRevE.51.4626,PhysRevE.52.4134} and is based on the Lennard--Jones potential
\begin{equation}
U_{ab}(r)
=4F_{ab}\left[\left(\frac{E_{ab}}{r}\right)^{12}
-\left(\frac{E_{ab}}{r}\right)^{6}\right],
\label{lj}
\end{equation}
where $a,b\in\{A,B\}$. The interaction parameters are
$F_{AA} = 1.0$, $E_{AA} = 1.0$,
$F_{AB}= 1.5$, $E_{AB} = 0.8$,
$F_{BB} = 0.5$, and $E_{BB} = 0.88$.
The model consists of A and B particles with equal mass $m$ in a 65{:}35 proportion.

We use reduced units: length is scaled by $E_{AA}$, temperature by $F_{AA}/k_{\mathrm{B}}$ (with $k_{\mathrm{B}}=1$), and time by $(mE_{AA}^{2}/F_{AA})^{1/2}$.  
The system contains 4000 particles under periodic boundary conditions.  
In the NVT ensemble, the density is fixed at $1.162$ and the temperature is cooled from $2.1$ to $0.1$ at a rate of $1\times10^{-5}$, using an integration timestep of $0.005$.  
The glass transition temperature of the model is approximately $0.33$ \cite{bruning:hal-00365282}. Kinetic quantities are computed in a standard microcanonical (NVE) ensemble.  
All molecular dynamics simulations were performed using LAMMPS \cite{THOMPSON2022108171}.

\subsection{Mode analysis}
\begin{figure}[ht]
    \centering
    \includegraphics[width=\linewidth]{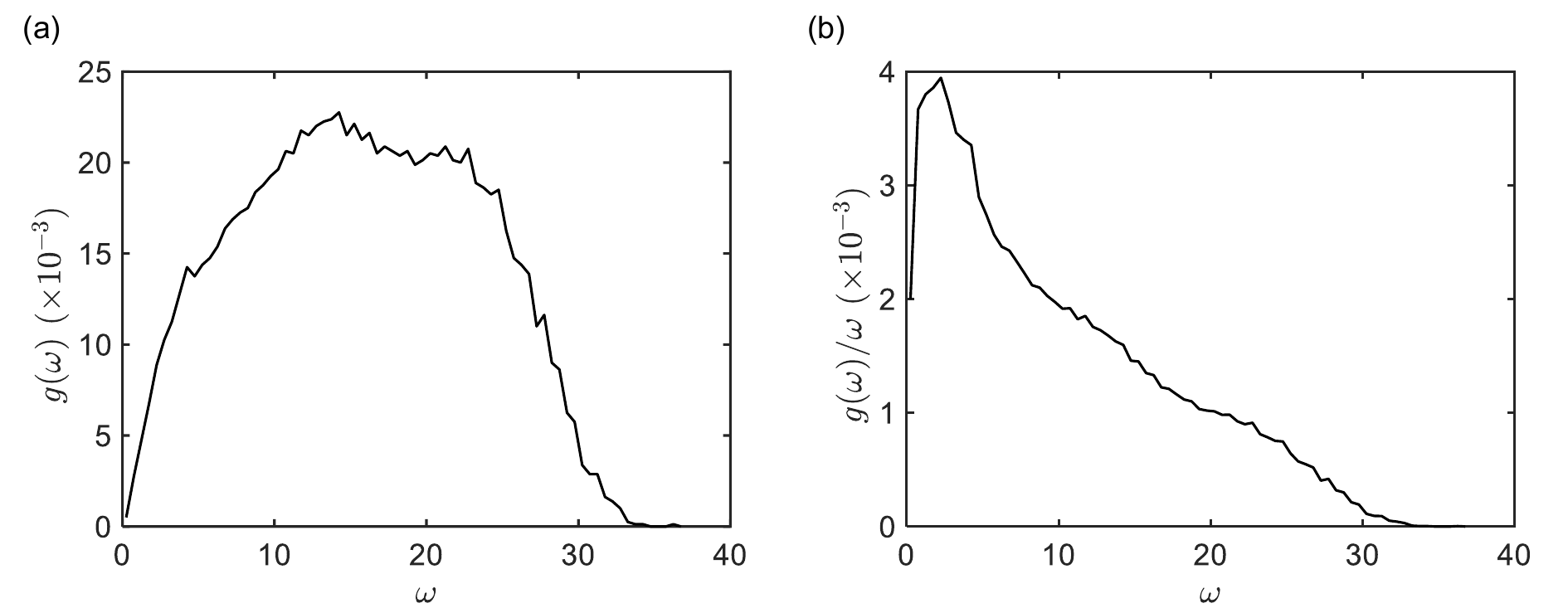}
    \caption{\textbf{(a)-(b)} Vibrational density of state (DOS) and reduced DOS of 2D Kob-Andersen glass respectively.}
    \label{fig:s1}
\end{figure}
The inherent structure at low temperature is obtained by minimizing the potential energy using the Fast Inertial Relaxation Engine (FIRE) algorithm \cite{PhysRevLett.97.170201}.  
The dynamical matrix of the inherent configuration is then diagonalized to obtain all eigenfrequencies and eigenvectors. The resulting vibrational density of states is presented in Fig. \ref{fig:s1}(a). A boson peak excess is clearly visible in the Debye reduce VDOS shown in panel (b) of the same figure.

\subsection{AQS protocol}

The inherent configuration is subjected to athermal quasi-static shear (AQS) \cite{PhysRevE.74.016118} in both the $xy$ and $-xy$ directions, using shear increments of $\Delta \gamma = 1\times10^{-5}$ per step. The corresponding stress-strain curve is shown in Fig.~\ref{fig:4}(a) in the main text.

\subsection{Numerical method}
\begin{figure}[ht]
    \centering
    \includegraphics[width=0.5\linewidth]{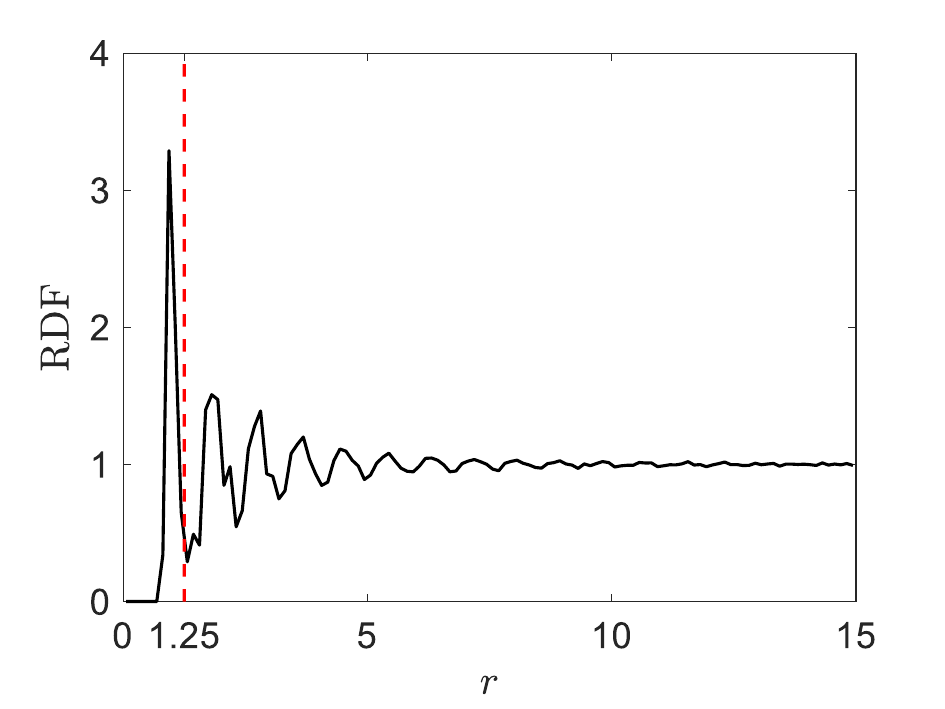}
    \caption{Radial distribution function (RDF) of 2D Kob-Andersen glass inherent structure. The red dotted line indicates the first minimum of the RDF $d_{c}=1.25$.}
    \label{fig:s2}
\end{figure}

\begin{figure}[ht]
    \centering
    \includegraphics[width=\linewidth]{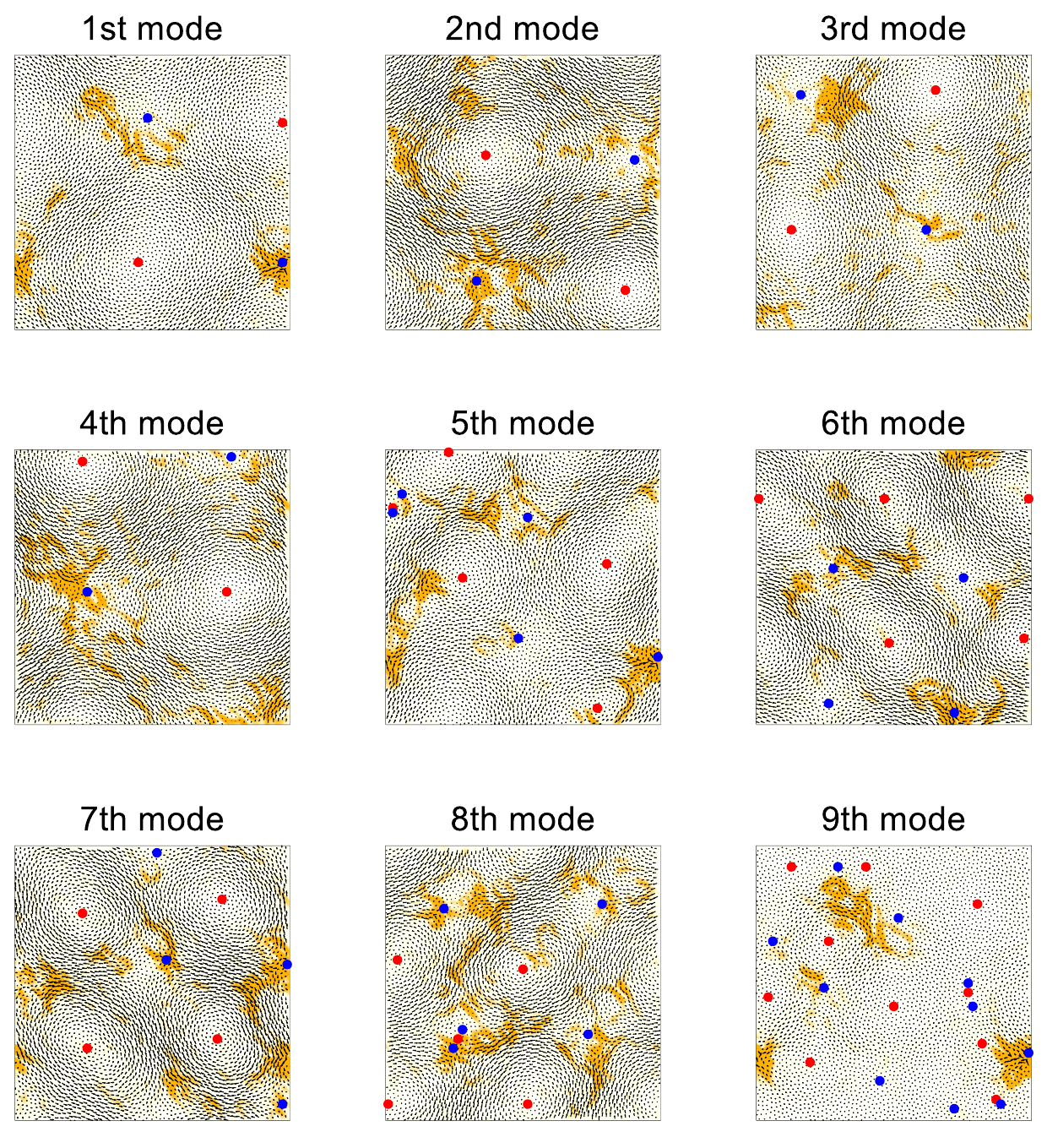}
    \caption{The eigenvector fields, topological defects and the local Nye density of the first nine modes; \bluec{3pt} and \redc{3pt} denote the $-1$ and $+1$ topological vortex defects. In all figures, the range of background color value is $[0,\mu(\rho) + 2\sigma(\rho)]$.}
    \label{fig:s3}
\end{figure}

The 2D simulation cell is divided into a $60\times60$ grid with sizes of grid d$r=1$.  
A smooth representation of the eigenvector field at each grid point $\vec{r}$ is constructed via a Gaussian weighting:
\begin{equation}
    \vec{e}^{\,\omega}(\vec{r})
    = \sum_i \vec{e}^{\,\omega}(x_i,y_i)\,
    \exp\!\left[-\frac{|\vec{r}-\vec{r}_i|^2}{r_c^2}\right],
\end{equation}
where $\vec{r}_i$ is the position of particle $i$ and $r_c=1$.  
The winding number $q$ and Nye density $\rho$ are then computed using Eqs.~\eqref{eq1} and \eqref{nye}, respectively. The topological vortex defects and the local Nye density of the first nine modes are shown in Fig \ref{fig:s3}.

\subsection{Nye-density clusters}
\begin{figure}[ht]
    \centering
    \includegraphics[width=\linewidth]{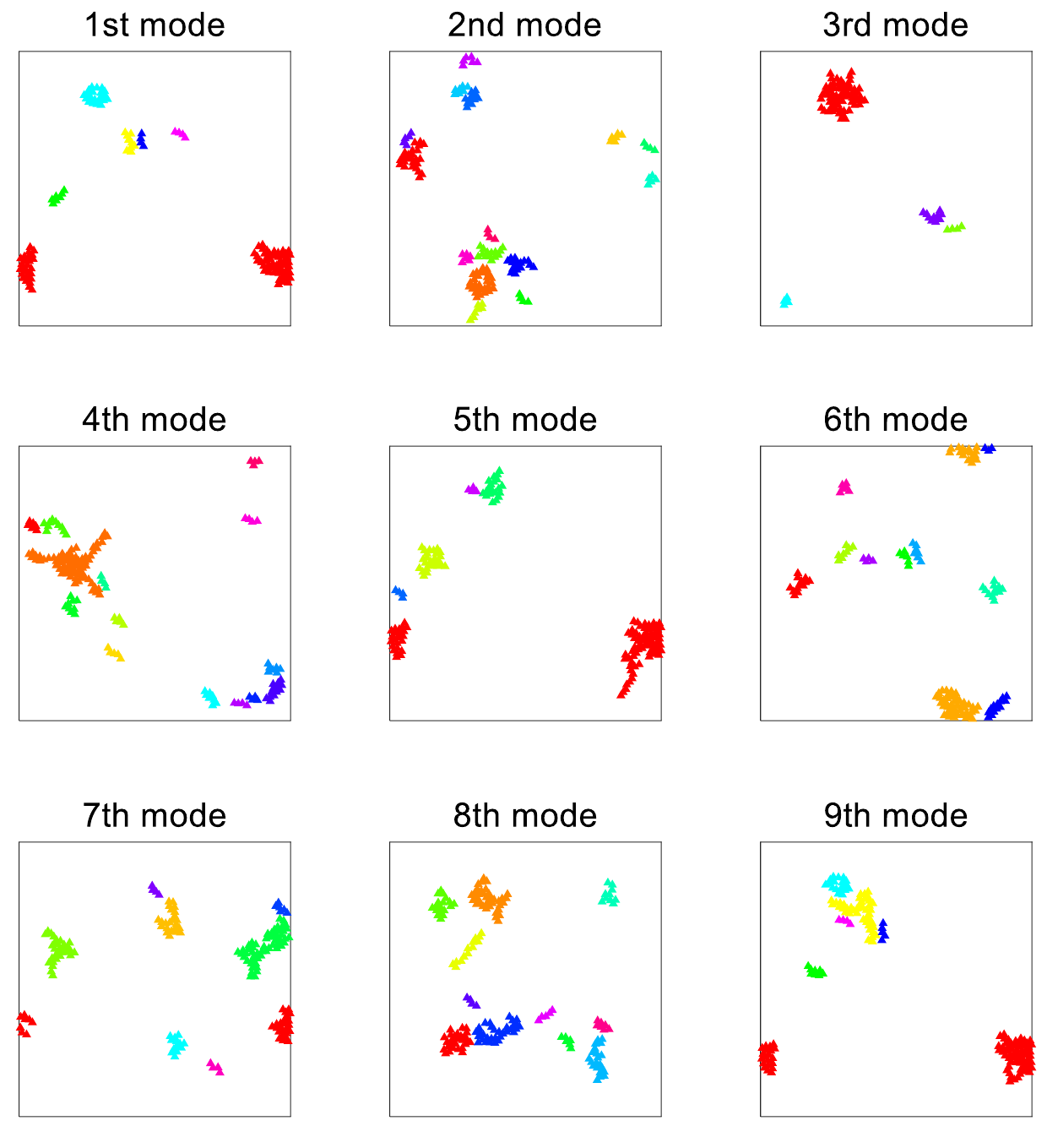}
    \caption{The Nye-density clusters of the first nine modes; different colors represent different clusters.}
    \label{fig:s4}
\end{figure}
To quantify the defects identified by the Nye density $\rho$, we first interpolate the grid-based $\rho$ field back onto atomic positions using biharmonic spline interpolation.  
Atoms with $\rho > \mu(\rho) + 2\sigma(\rho)$ (where $\mu$ and $\sigma$ are the mean and standard deviation) are selected.  
Two such atoms are assigned to the same cluster if their mutual distance is smaller than the position of the first minimum of the radial distribution function (RDF) $d_{c}=1.25$, as shown in Fig. \ref{fig:s2}.  
Clusters containing at least four atoms are retained. Figure \ref{fig:s4} shows the clusters in the first nine modes.

It is interesting to note that $-1$ anti-vortex defects (\bluec{3pt}) are often associated with regions where the Nye density is large. In contrast, positive vortex defects (\redc{3pt}) predominantly appear in regions where the Nye density vanishes. This observation supports the more prominent role of $-1$ defects advocated in \cite{wu2023topology}.

\subsection{Correlation function}

The pair correlation function between quantities $X$ and $Y$ at frequency $\omega$ is defined as
\begin{equation}
    g_{X,Y}^{\omega}(r)= 
    \frac{S}{2\pi r\,\Delta r\, N_X N_Y}
    \sum_{i=1}^{N_X}\sum_{j=1}^{N_Y}
    \delta\!\left(r - |\vec{r}_{ij}|\right),
\end{equation}
where $X,Y\in\{\mathrm{P},\alpha,-1,+1\}$,
$S$ is the simulation area,
$N_{X,Y}$ are the number of plastic spots, cluster atoms, or topological defects, and  
$\vec{r}_{ij}$ is the distance between entities of type $X$ and $Y$.

A frequency-averaged, weighted correlation function is defined as
\begin{equation}
    g_{\mathrm{P},Y}(r)
    = \sum_{i=1}^N
    \frac{g^{\omega_i}_{\mathrm{P},Y}(r)/\omega_i^2}
    {\sum_{i=1}^N 1/\omega_i^2},
\end{equation}
where $N$ is the number of modes included in the average.

\subsection*{Stability of the numerical method}
\begin{figure}[ht]
    \centering
    \includegraphics[width=\linewidth]{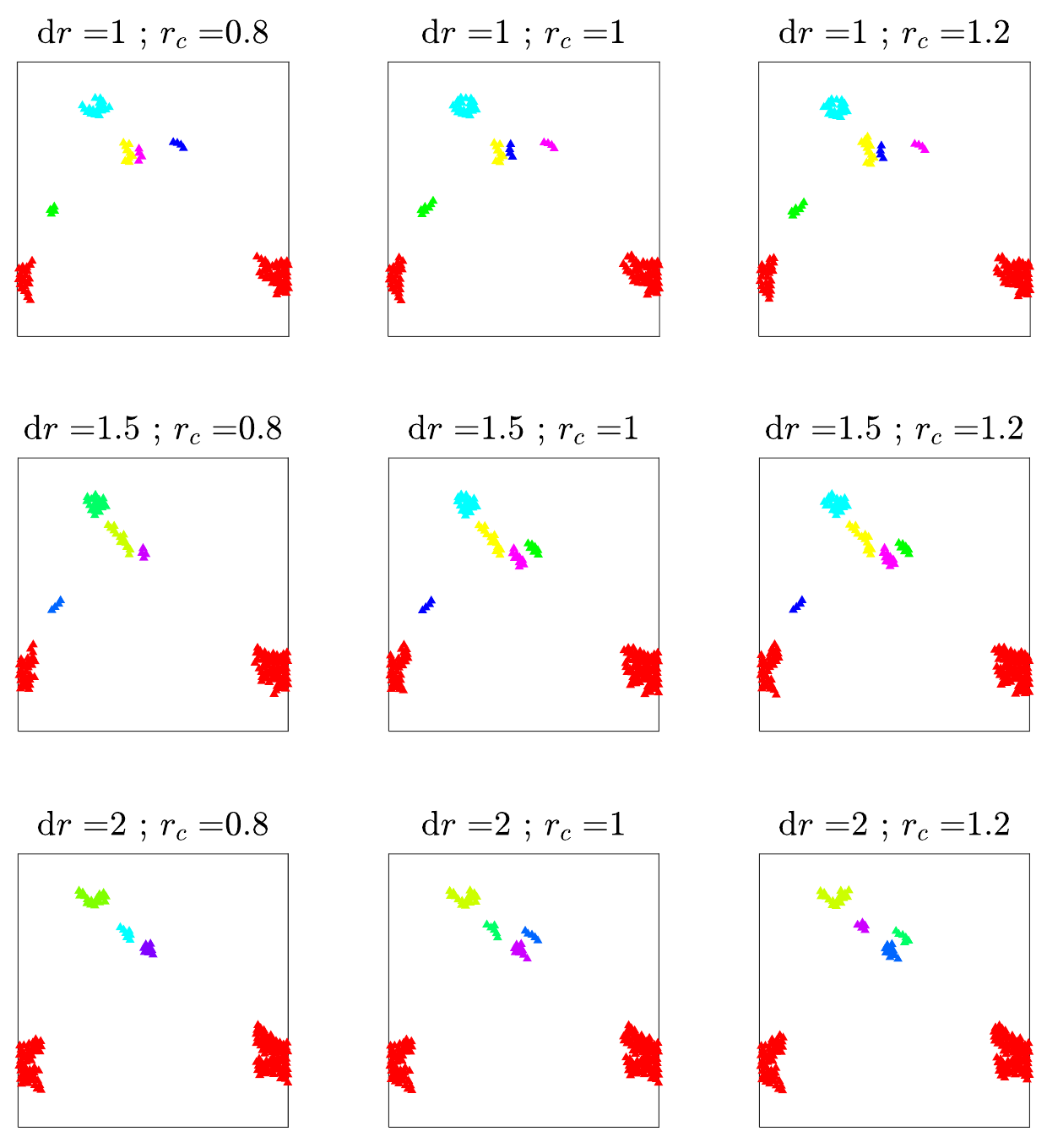}
    \caption{The Nye-density clusters of the first mode with different sizes of grid d$r$ and cutoff $r_c$; different colors represent different clusters.}
    \label{fig:s5}
\end{figure}

We vary the grid spacing $\mathrm{d}r$ and the cutoff radius $r_c$ to compute the Nye-density clusters for the first mode and assess the robustness of the algorithm, as shown in Fig.~\ref{fig:s5}. The resulting clusters remain stable across all parameter choices, confirming the reliability of our numerical procedure.

\subsection*{Different choices of the weight function}
\begin{figure}[ht]
    \centering
    \includegraphics[width=0.8\linewidth]{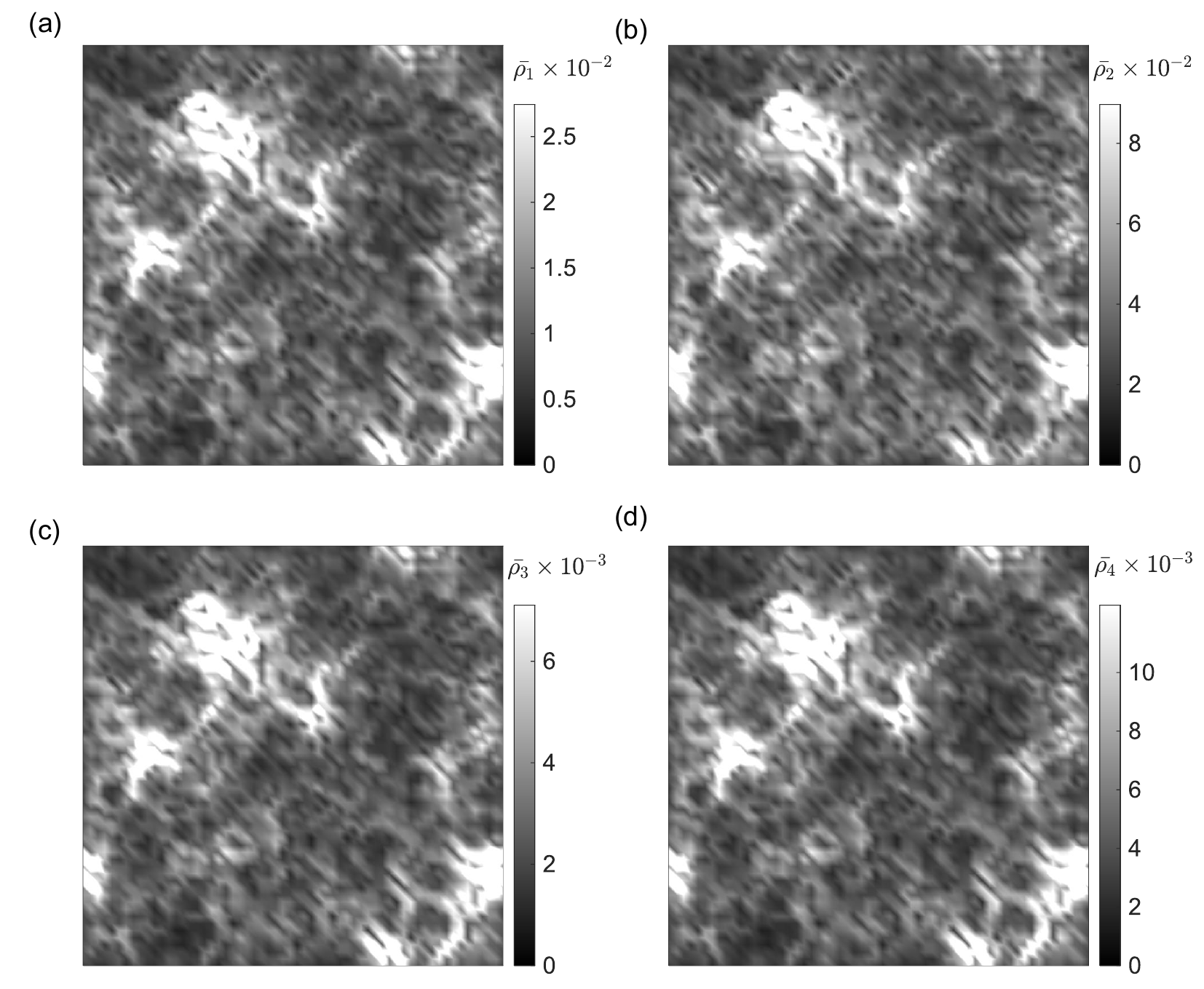}
    \caption{We compare different weighting schemes, Eqs.~\ref{ww1}-\ref{ww4}, for computing the average Nye density over the first $n=20$ modes. The similarity between the different schemes is evident.}
    \label{fig:s6}
\end{figure}

Here, we compare different weighted methods to present the average value of the Nye density:
\begin{equation}
    \bar{\rho_1}\equiv \frac{1}{n}\sum_{i=1}^n \frac{\rho_i(x,y)}{\omega_i^2},\label{ww1}
\end{equation}

\begin{equation}
    \bar{\rho_2}\equiv \frac{1}{n}\sum_{i=1}^n \frac{\rho_i(x,y)}{\omega_i^4},\label{ww2}
\end{equation}

\begin{equation}
    \bar{\rho_3}\equiv \frac{1}{n}\sum_{i=1}^n \rho_i(x,y)\exp(-\omega_i^2),\label{ww3}
\end{equation}

\begin{equation}
    \bar{\rho_4}\equiv \frac{1}{n}\sum_{i=1}^n \rho_i(x,y),\label{ww4}
\end{equation}
As shown in Fig.~\ref{fig:s6}, the average Nye density computed using different weighting schemes over the first $n=20$ modes yields essentially the same result.

\end{document}